%% file: main.tex
\documentclass[12pt]{article}
\usepackage{graphicx} 
\usepackage{listings}
\usepackage{amsmath} 
\usepackage{hyperref}
\usepackage{multirow}
\usepackage{amsfonts}
\usepackage{lscape}
\usepackage{palatino} 

\usepackage{float}
\usepackage{subcaption}
\usepackage{booktabs}

\usepackage{pdflscape}
\usepackage{tikz}
\usetikzlibrary{positioning}    
\usepackage[doublespacing]{setspace}
\hypersetup{
	colorlinks=true,
	citecolor=blue,
	pdfborder={0 0 0},
}
\usepackage{diagbox}
\usepackage{makecell}

\usepackage{mathrsfs}
\usepackage{enumitem}
\usepackage{amssymb}
\usepackage[margin=1in]{geometry}
\usepackage{indentfirst}
\title{\huge  Measuring Regulatory Barriers Using Annual Reports of Firms}
\author{Haosen Ge\thanks{Ph.D. Candidate, Department of Politics, Princeton University, \href{mailto:hge@princeton.edu}{hge@princeton.edu} }}
\date{}

\usepackage[title]{appendix}

\usepackage{ltablex,caption}

\usepackage{rotating, lscape, threeparttable}

\usepackage[
backend=biber,
style=authoryear,
uniquename=false
]{biblatex}
\bibliography{ref.bib}

\usepackage{cellspace}
\begin{document}

\begin{titlepage}

\maketitle

\begin{abstract}
	
\textit{Existing studies show that regulation is a major barrier to global economic integration. Nonetheless, identifying and measuring regulatory barriers remains a challenging task for scholars. I propose a novel approach to quantify regulatory barriers at the country-year level. Utilizing information from annual reports of publicly listed companies in the U.S., I identify regulatory barriers business practitioners encounter. The barrier information is first extracted from the text documents by a cutting-edge neural language model trained on a hand-coded training set. Then, I feed the extracted barrier information into a dynamic item response theory model to estimate the numerical barrier level of 40 countries between 2006 and 2015 while controlling for various channels of confounding. I argue that the results returned by this approach should be less likely to be contaminated by major confounders such as international politics. Thus, they is well-suited for future political science research.}
\end{abstract}
\end{titlepage}

\newpage

\section{Introduction}

Regulations have drawn increasing attention from policymakers in recent years. The World Trade Organization (WTO) identifies regulatory barriers as the most salient obstacle to economic globalization in the past decade.\footnote{World Investment Report 2008, 2018} Researchers show that regulations have become a powerful tool wielded by multinational firms to protect their special interests \autocite{buthe_new_2011,carpenter2013preventing,gulotty_narrowing_2018,perlman2019,kennard2017firms}. Nonetheless, due to the complexity of regulatory regimes, understanding the causes and effects of regulatory barriers often turn out to be very challenging. The obstacle is especially pronounced when scholars attempt to study regulatory barriers systematically. 
Regulations vary considerably by country. It is generally infeasible to assess how stringent regulations are by studying their texts, as understanding texts of regulations requires in-depth knowledge of the operation of an industry in a specific country. To make things worse, the distributional effects of regulations are often so subtle that researchers cannot tell who the winners and losers are by only reading the texts. For example, rules on automobile parts in a country are applied to all auto manufacturers, domestic or foreign. Yet, the impact of such regulations is far from uniform across auto manufacturers. To a large extent, the difficulty of identifying and measuring regulatory barriers has limited the research progress in that area. In this paper, I propose a novel approach to measuring regulatory barriers at the country-year level. Leveraging on the information included in the annual reports of U.S. public companies, I build an original database of observed regulatory barriers and use it to estimate the barrier level of each country with a dynamic two-level item response theory model. This new approach helps us reveal new empirical patterns that are previously unobservable to scholars.

My main source of information comes from the United States Securities and Exchange Commission (SEC). SEC requires all U.S. public firms to disclose information about their operation in their annual reports (i.e., 10-K forms). The annual reports submitted to SEC are much more detailed than the annual reports published to share information with the firms' shareholders
\footnote{\href{https://www.sec.gov/fast-answers/answers-form10khtm.html}{Detailed explanations on 10-Ks}}. 
Specifically, federal laws require every public firm to list all major factors that may adversely impact their performance in their 10-K forms. For that reason, companies that are subject to regulatory barriers are required to report their encountered difficulties. This valuable information provides us with the knowledge that we otherwise would not know: whether a firm is adversely affected by specific regulations. As explained by \textcite{kennard2017firms}, regulations do not evenly affect all firms in an industry; instead, it is usually the case that rules can tilt the playground in favor of some firms but against others. It is almost impossible for scholars to tell how the ``playground'' is tilted without acquiring any information from the ``players'' (i.e., firms). Thus, the information on the adverse impacts of regulations included in the firms' 10-K forms can be valuable for researchers. 

To quantify the level of regulatory barriers for each country, I first convert the information in these reports into a well-structured dataset using text processing techniques. Then, I build a dynamic item response theory model to estimate the level of regulatory barriers at the country-year level. I believe that this new approach contributes to our existing knowledge in the following aspects:
\begin{enumerate}
	\item Compared with traditional survey-based measurement, my approach can be easily extended to include more countries for a longer time period. More importantly, it can estimate \textit{the entry-deterrence effects of regulations} under reasonable assumptions.  \textcite{gulotty_narrowing_2018} argues that one of the major effects of regulations is entry-deterrence: large firms advocate more stringent industry standards to forestall new firms from entering by raising the fixed cost of operation. For example, increasing the quality standard of a product will increase the cost of production, which can eliminate small producers and deter other small firms from entering the market. However, existing survey-based measurements only survey firms that are active in the market and hence cannot estimate the entry-deterrence effects. In addition, due to the cost of running surveys, the coverage of existing survey measurements is far from ideal. For example, the NTM Business Surveys run by the International Trade Commission only cover around 25 countries which do not include large countries such as the United States, China, Japan, Germany, South Korea, India, and many others.\footnote{They claim to have surveyed the EU, but I cannot find the data anywhere on their website at the time of writing this chapter}
	\item The proposed approach is more micro-founded than data collected by international organizations, such as the Special Trade Concerns (STCs) Report data compiled by WTO. Members/Observers of WTO can raise STCs against other countries if they find laws and regulations in other countries discriminatory. Scholars use whether a country is subject to STCs to measure whether the country imposes any regulatory barriers. However, it is well documented that the behaviors of countries in WTO are shaped by political concerns \autocite{davis2012adjudicate}. The STCs records are very likely to be outcomes of both domestic regulations and international politics. It is then problematic if scholars of political science use it to study countries' political behaviors. However, the firm-level information in the annual reports is less susceptible to international and domestic political factors. For that reason, my proposed measurement should be better suited for studies in political science.
	\item  Existing work has shown promising progress in estimating non-tariff barriers to trade (e.g., \textcite{cooley2019barrier,martini2020backward}). The proposed approach complements their contributions by adding information about barriers to foreign direct investment and operations of foreign subsidiaries. 
\end{enumerate}

\section{Information in 10-K Forms}

The U.S. federal securities laws require public companies to disclose information on an ongoing basis. All U.S. public firms must submit annual reports, a.k.a. form 10-K, which provides a comprehensive overview of the company's business and financial condition. In the 10-K forms, a company must offer a detailed description of its main products or services, major subsidiaries, relevant regulations, major competition, and any possible risks associated with its business. Therefore, the 10-K forms contain information on a comprehensive set of regulatory barriers observed by international business practitioners.

 To better illustrate the type of information included in 10-K forms, I will present some examples here. Many firms report that they have encountered restrictive laws or regulations in certain countries:
\begin{enumerate}
	\item A wholesale drug company Nu Skin Enterprises reports in their 2006 10-K form: ``laws and regulations in Japan, Korea and China are particularly restrictive and difficult.''
	\item A farm machinery producer Deere \& Co reports that: ``recent industry and regulatory changes have negatively impacted John Deere's competitive position in the potential high growth Russian markets during the fiscal year.''
	\item An ophthalmic goods producer Cooper Companies INC claims that: ``we have difficulty gaining market share in countries such as Japan because of regulatory restrictions and customer preferences.''
	\item A medical equipment producer Immucor Inc states that: ``in addition to the U.S., Europe, Canada and Japan, there are multiple countries worldwide that also impose regulatory barrier to market entry.''
	\item  An insurance company Gerova Financial Group Ltd claims that ``The Chinese and Vietnamese governments have imposed regulations in various industries, including the leisure and hospitality and financial services industries, that would limit foreign investors  equity ownership or prohibit foreign investments altogether in companies that operate in such industries.''
	\item A software company Versant Corp reports that they are faced with ``burdens of complying with a variety of foreign laws, including more protective employment laws affecting our sizable workforce in Germany''.
	\item  A technology company Kenexa Corp reports their concerns about the intellectual property issues: ``Further, the laws of some countries, and in particular India, where we develop much of our intellectual property, do not protect proprietary rights to the same extent as the laws of the United States.''
	\item A technology company National Instrument Corp reports the difficulties of doing business in Hungary: ``In response to significant and frequent changes in the corporate tax law, the unstable political environment, a restrictive labor code, the volatility of the Hungarian forint relative to the U.S. dollar and increasing labor costs, we have doubts as to the long term viability of Hungary as a location for our manufacturing and warehousing operations.''
\end{enumerate}

It is evident from these examples that reporting regulatory barriers in 10-K forms is a common practice among firms. However, one may question whether reporting the existence of regulations is a good indicator of the firm being adversely affected by them. In other words, the presence of regulations does not necessarily imply whether they constitute barriers for firms. To address this concern, I present three more examples to show that the information on regulations included in the annual reports is in fact related to the distributional effects of regulations.

\begin{enumerate}
	\item In 2005, China passes a regulation that mandates all truck manufacturers to install electrical throttle to reduce emission. This piece of regulation obviously affects all major truck producers serving the Chinese market. Cummins, a U.S. based natural gas engines producer who has joint ventures in China, reports in their 10-K form that ``These (earning) increases were partially offset by decreased earnings from DCEC (one of Cummins' joint venture in China) of \$7 million due to reduced demand in China's truck market in response to regulatory changes.'' It is quite evident that the operation of Cummins is harmed by the newly implemented regulation. However, in the same year, another U.S. based automobile parts manufacturers, Williams Control, found the new regulation an opportunity rather than a hurdle. In their 10-K form, they reported: ``Increases in off-road volumes in China primarily results from adoption of more stringent emissions standards, which mandate the inclusion of electronic throttle controls on new vehicles, thus allowing us to expand our customer base in this market.'' We can immediately tell from this two pieces of information that Cummins is the loser of the new regulation while Williams Control is the winner.
	\item Still in 2005, a cosmetic and fragrance producer Inter Parfums reports that the existing regulations in France have little effect on their operation: ``our fragrances that are manufactured in France are subject to certain regulatory requirements of the European Union, but as of the date of this report, we have not experienced any material difficulties in complying with such requirements.''
\end{enumerate}

The cases presented above demonstrate what these firms report is closely related to how they are affected by the reported regulations. If a firm finds a law positively impacts their business, they will have strong incentives to report it truthfully since their 10-K forms are made available to all shareholders. Meanwhile, if a firm encounters a harmful regulatory barrier, it may or may not want to share it with the public, but federal laws make it mandatory for them to disclose that piece of information. For these reasons, I argue that firms' annual reports are both informative and truthful, which makes them ideal sources for studies of regulatory politics.

\subsection{The Truthfulness of 10-K Forms}

I will further justify the 10-Ks truthfulness in the subsection, as the plausibility of my proposed barrier index crucially depends on the information accuracy.

Since 10-K forms can significantly impact a firm's stock market performance, one may be suspicious about the authenticity of its information. The scrutiny is especially warranted when we evaluate how firms report their encountered regulatory barrier, due to the inherent ambiguity of regulations. In this section, I would like to establish the reliability of 10-Ks by providing more substantive information on how public companies write these reports and how SEC enforces the disclosure requirements. Hopefully, readers can be more assured about using 10-Ks as a data source.

Public firms have two main concerns when writing the annual reports: 1) they are concerned with the shareholders instituting legal actions against them for financial loss resulting from undisclosed issues, and 2) the punishment from SEC if found guilty of hiding information. For example, In 2013, stockholders sued a company called Dole Foods for failure to disclose positive information. In the end, the judge found the company guilty of unfairly keeping the stock price down.\footnote{See \href{https://www.edgarlawfirm.com/blog/2019/08/two-reasons-corporation-shareholders-sue/}{a report}} In this monitoring process, the shareholders serve as a ``fire alarm'' that forces the public firms to disclose any positive and negative information honestly \autocite{mccubbins1984congressional}. 

In addition to shareholders' monitoring efforts, the SEC also actively takes measures to ensure the company's annual reports' authenticity. After firms submit their 10-K forms, the SEC staff will review the submissions to monitor and enhance companies' compliance with the requirements. If the review process finds the disclosed information deficient in explanation or clarity, the SEC staff will provide comments for a company to resolve the issues. Moreover, the SEC has made disclosure of qualitative information (such as the risk factor subsection) a focus of its corporate filing reviews \autocite{campbell2014information,brown2018spillover}. From 2013 to 2017, the chairman, Mary Jo White emphasized the effectiveness of information disclosure which urged firms to include more relevant qualitative information in their disclosures. 

In reality, the SEC also imposes significant punishment on firms that fail to disclose crucial information. Facebook, for example, is expected to pay \$100 million for making misleading disclosures regarding their user privacy policy.\footnote{Facebook to Pay \$100 Million for Misleading Investors About the Risks It Faced From Misuse of User Data: \href{https://www.sec.gov/news/press-release/2019-140}{link}} The SEC's complaint alleges that Facebook failed to disclose its risk of a data breach even after it had discovered the misuse of its users' information in 2015.

The above evidence suggests that the SEC is well aware of the issue of dishonest reporting and has taken a series of actions to enforce the filing requirement. According to a study by \textcite{brown2018spillover}, firms improve their filling quality after the SEC issues comments to them during the review process. 

For these reasons, public firms usually utilize two strategies to minimize the legal risks: 1) they timely add issues that might adversely affect their operation in the annual reports, and 2) once added, the issues are rarely removed from a future version of the reports. I argue that these two features make the annual reports ideal for measuring regulatory barriers. First, it is very costly for public firms to misreport in the annual reports. Thus, as researchers, we should be confident in the information quality. Second, the reluctance of firms to remove negative issues from the report makes it easier to interpret the reported barrier. Barrier estimates based on the reported barrier information can be viewed as cumulative regulatory barriers instead of the change in regulatory barrier levels. As a result, the 10-K forms can be beneficial for international political economy research.

\section{Text Processing}

To quantify the information in the annual reports, I need to identify instances where a firm reports being adversely affected by a specific piece of regulation. The text in annual reports is not well-structured: different firms have different reporting formats. In addition, the annual reports often contain several hundred pages of text filled with technical terms. Thus, converting the annual reports to a well-structured dataset is challenging.

In this paper, I adopted the following strategy to create a feasible data processing pipeline, which combined a dictionary-based method with supervised learning:
\begin{enumerate}
	\item First, I break each report into sentences by using a regular expression.
	\item Second, I select sentences using a dictionary of regulation-related words: ``regulation, regulator, regulatory, law, standard, quota, approval, policy, intellectual property, requirement, permit, license''. In addition, I filter out sentences that do not contain a country name.
	\item Next, I randomly sample 3,846 sentences for human coding, with the help of two research assistants. If the sentence clearly indicates that the firm is adversely affected by a regulation, the sentence is coded as ``1'' (i.e., positive). Otherwise, the sentence is coded as ``0'' (i.e., negative).
	\item  Finally, I train a neural network model to predict the rest of the corpus. If a report contains any sentence the model predicts to be positive, I code the entire document as the firm reporting barriers in that country.
\end{enumerate}

This strategy is a balance between accuracy and feasibility. Annotating annual reports with human coders is almost impossible, as the reports are too long and highly technical. However, the task is greatly simplified if we only code specific sentences instead of the entire report. Admittedly, filtering the document with a dictionary of regulation-related words introduces (unmeasurable) errors. We leave improving the text processing pipelines to future research.

Training a prediction model capable of dealing with complex sentence structures with a limited sample size is the next challenging task. In this paper, I choose a state-of-the-art neural language model, ``BERT'', to perform this task. ``BERT'' is a neural language model pre-trained by scientists at Google \autocite{devlin2018bert}. It is designed to understand the context of a sentence and predict the appropriate words suitable for the context. The model is trained on a vast corpus that includes the entire Wikipedia (2,500 million words) and BookCorpus (800 million words. As a result, ``'BERT' is a powerful tool for most text classification tasks.

I train a classifier on my sample of 3,846 sentences using ``BERT'' as the underlying workhorse model (i.e., fine-tuning). The training sample includes 3,486 sentences, while I leave the other 400 sentences as the test set. After training, the model returns satisfactory prediction accuracy: among the 400 test sentences, the results are:
\begin{table}[h!]
	\caption{Model Performance: Confusion Matrix}
	\label{confuse}
	\centering
	\begin{tabular}{ccc}
		\toprule
		& Actual True & Actual False \\ \midrule
		Predicted True      & 44          & 48           \\ \midrule
		Predicted False     & 6           & 302          \\ \midrule
		False Positive Rate & \multicolumn{2}{c}{0.12}  \\ \midrule
		False Negative Rate & \multicolumn{2}{c}{0.137} \\ \midrule
		Total Error Rate    & \multicolumn{2}{c}{0.135} \\ \bottomrule
	\end{tabular}
\end{table}

Since the training set has very few positive examples (i.e., only a few sentences contain information on regulatory barriers), monitoring both the false positive and negative rates is crucial. In the sparse prediction task, a model that blindly returns ``0'' can still achieve seemingly perfect accuracy simply because the number of positive examples accounts for a tiny proportion of the training set. Fortunately, my model performs well in distinguishing positive and negative examples, illustrated by the low false negative rate. As shown in \hyperref[confuse]{Table \ref{confuse}}, both the false positive rate and the false negative rate are less than 15\%, demonstrating that the model refrains from blindly assigning ``0''.

Using the fine-tuned BERT as the machine reader, I assign 0 or 1 to all sentences in the corpus. If any sentence in an annual report is predicted to be positive, I classify the example as the firm reporting regulatory barriers. Then, I extract the country names by comparing the text with the list of all country names. The final data contains a matrix of $I$ firms and $N$ countries over $T$ years. The problem, however, is to aggregate information from the firm-country-year triad level to the country-year dyad level, as we need to compare regulatory barrier levels both across country and across time. A naive approach is to take the mean reporting level at the country-year level, but such an approach requires strong assumptions that are unsubstantiated by theories. Therefore, I use a widely-accepted latent factor model to estimate regulatory barriers, a similar solution to that used by \textcite{hollyer2014measuring}.

\section{The Statistical Model}

\subsection{Setup}
The statistical model aims to assign a scalar-valued index to each country in each year while accounting for major firm and country heterogeneity. I propose to use an item response theory model to estimate this quantity of interest, a popular model often used in estimating ideological positions of legislators \autocite{clinton2004statistical} but has become increasingly popular in studies of international relations (e.g., \textcite{hollyer2014measuring,bailey2017estimating}).

First, I convert the firm reported barrier into a well-structured dataset (\hyperref[ds]{Table \ref{ds}}). In a given year $t$, we have a $I \times J$ matrix $\{U_{ijt}\}$ with $i \in \{1, 2, 3, \cdots, I\}$ and $j \in \{1, 2, 3, \cdots, J\}$, where $I$ indexes the total number of firms in the sample and $J$ indexes the number of countries:

\begin{table}[h!]
	\centering
	\caption{Data Structure}
	\label{ds}
	\begin{tabular}{@{\extracolsep{5pt}}ccccc}
		\\[-1.8ex]\toprule
		& Country 1 & Country 2 & $\cdots$ & Country $J$\\
		\midrule
		Firm 1& $U_{11t}$ & $U_{12t}$ & $\cdots$ & $U_{1Jt}$ \\
		Firm 2 & $U_{21t}$ & $U_{22t}$ & $\cdots$ & $U_{2Jt}$ \\
		\vdots & \vdots & \vdots & \vdots& \vdots \\
		Firm $I$ & $U_{I1t}$  & $U_{I2t}$  & $\cdots$ & $U_{IJt}$\\
		\bottomrule
	\end{tabular}
\end{table}

The term $U_{ijt}$ can take three possible values:
\begin{align*}
U_{ijt} =
\begin{cases}
3 & \text{firm $i$ does not enter country $j$ in year $t$}\\
2 & \text{firm $i$ enters country $j$ AND reports barrier in year $t$}\\
1 & \text{firm $i$ enters country $j$ AND does not reports barrier in year $t$}
\end{cases}
\end{align*}

I assume that each country $j$ has a regulatory barrier level $\theta_{jt}$ in year $t$ that is observable to firms but is not to researchers. Each firm-country dyad $(i,j)$ in year $t$ has two dyadic characteristics: an entry cutoff $\alpha^E_{ij}$ and a reporting cutoff $\alpha^R_{ij}$, for which I will provide a more detailed exposition later. The observed firm behavior $U_{ijt}$ is a function of the country barrier level $\theta_{jt}$ and the dyad-specific reporting and entry cutoffs. Note that the cutoffs are time-invariant, a strong assumption to avoid the model identification problems.

The intuition behind such a setup is straightforward. Each firm $i$ observes the regulatory barrier level of country $j$ in year $t$ (i.e., $\theta_{jt}$) and chooses an action. The reporting and entry cutoffs capture the firms' tolerance level of a country's barrier. The concept of tolerance is an abstraction from firms' real-world calculations, which often include market size, cultural similarity, and geographical distance. Because firms' decisions are affected by a multitude of factors, I wrap any factors that shape firms' decisions but are not components of a country's regulatory barriers into the tolerance terms, to alleviate concerns of omitted variables bias caused by firm and country heterogeneity.

Specifically, if a country's barrier level is higher than a firm's entry tolerance cutoff (i.e., $\alpha^R_{ij}$), the firm will not operate in that country; if, on the other hand, the country's barrier level is below the firms' entry tolerance cutoff but higher than the reporting tolerance cutoff, the firm will enter the country but report the barrier in their annual reports; lastly, if the country's barrier level is lower than both the entry and reporting cutoff level, the firm will operate in that country and no barrier reporting will be witnessed in the annual reports. To fix ideas, I will provide a formal explanation of this logic later in this section.

For ease of exposition (and model estimation), I define a random latent firm-country barrier level $U_{ijt}^*$:
\begin{align*}
	U^*_{ijt} = \theta_{j,t} + \epsilon_{ijt}
\end{align*}
That is, each firm $i$ perceives the barrier level of country $j$ in year $t$ as slightly different. The random disturbance $\epsilon_{ijt}$ captures any idiosyncrasies at the firm-country-year level. Readers can interpret the errors as factors that are unobservable to researchers but affect firms' perception of countries' regulatory barriers. Following the practice in the scaling literature, I assume that the disturbance term follows a standard normal distribution $\mathcal{N}(0, 1)$.

Each firm $i$ compares the latent firm-country barrier level $U_{ijt}^*$ with its entry and reporting cutoffs and chooses an action according to the following decision rule:
\begin{itemize}
	\item If country $j$'s latent barrier level is higher than firm-country $(i.j)$'s dyadic entry cutoff (i.e., $U^*_{ijt} > \alpha_{ij}^E$), firm $i$ will not enter country $j$ (i.e., $U_{ijt} = 3$). 
	\item  If country $j$'s barrier level is lower than $(i,j)$'s entry cutoff but the barrier level is higher than the dyad's reporting cutoff (i.e., $  \alpha_{ij}^E > U^*_{ijt} > \alpha^R_{ij}$, firm $i$ will enter country $j$ but report encountering  barrier (i.e., $U_{ijt} = 2$).
	\item If country $j$'s barrier level is lower than the dyad $(i,j)$'s reporting cutoff (i.e., $ U^*_{ijt} \leq \alpha_{ij}^R$), firm $i$ will enter country $j$ and not report any barrier  (i.e., $U_{ijt} = 1$).
\end{itemize}
Note that the implicit assumption is that the entry cutoff is greater than the reporting cutoff (i.e., $\alpha_{ij}^E > \alpha^R_{ij}$).

Since barrier level $U^*_{i,j,t}$ and $\theta_{jt}$ is unobservable to researchers, the proposed approach leverages the connection between regulatory barrier and firms' entry and report behaviors to make inference about the barrier level. It is worth noting that the model will interpret a firm not entering a country as a signal of prohibitive regulatory barriers after taking the firms' tolerance level into account. This is admittedly the strongest assumption of the model, as it fails to distinguish between a country being unattractive and a country imposing a significant entry barrier. I try to address this concern by only including the most globally active U.S. firms, which I will explain in detail in the next subsection. Nonetheless, readers should be cautious when interpreting the results.

This model can be translated into a statistical model by noting the relationship between the theoretical model and our observed data.
\begin{align*}
	U_{ijt} = \begin{cases}
		1 \quad \text{if} \quad U^*_{ijt} \leq \alpha_{ij}^R \\
		2  \quad \text{if}\quad  \alpha_{ij}^R  < U^*_{ijt} \leq  \alpha_{ij}^E \\
		3  \quad \text{if} \quad  U^*_{ijt} > \alpha_{ij}^E
	\end{cases}
\end{align*}

Denote the set of $\{\theta_{j,t}\}$ as $\Theta$ and the set of $\{\alpha_{ij}^E\}$ and $\{\alpha_{ij}^R\}$ as $\alpha^E$ and $\alpha^R$. Let $U$ denote the observed data and $U^*$ the augmented data. Then, we can write the full data likelihood with data augmentation as:
\begin{align*}
		\mathcal{L}(\Theta, \alpha^E, \alpha^R | U, U^*) = \prod_{t=1}^T \prod_{j = 1}^J \prod_{i = 1}^I [I(U_{ijt} = 1, U^*_{ijt} \leq  \alpha_{ij}^R) + I(U_{ijt} = 2,  \alpha_{ij}^R  < U^*_{ijt} \leq  \alpha_{ij}^E) \\
	+ I(U_{ijt} = 3, U^*_{ijt} > \alpha_{ij}^E)] \cdot \phi_{\theta_{jt}}(U_{ijt}^*)
\end{align*}
where $\phi_{\theta_{jt}}(\cdot)$ denotes the probability density function of $\mathcal{N}(\theta_{jt}, 1)$.

The high dimensionality of the model poses a considerable challenge for efficient estimation. Therefore, I propose a Gibbs sampler with a Kalman filter to achieve optimal estimation performance. Existing political science and economics studies have demonstrated the superiority of such an estimation strategy over a generic Gibbs sampler or a Metropolis-Hastings sampler \autocite{martin2002dynamic,west2006bayesian}, as it efficiently utilizes the time series information.

\subsection{Posterior}

Recall that the model aims to calculate the mean barrier level for each country in each year. We can obtain the desired quantity by calculating the mean of each parameter's posterior distributions.

First, priors distributions are necessary for calculating the posterior distributions,
\begin{align*}
	\alpha_{ij}^R|\alpha_{ij}^E &\sim \mathcal{N}_{(-\infty, \alpha_{ij}^E)}(0,5^2) \\
	\alpha_{ij}^E|\alpha_{ij}^R &\sim \mathcal{N}_{(\alpha_{ij}^R, \infty)}(0,5^2) \\
	\theta_{j,0} &\sim \mathcal{N}(0,5^2)
\end{align*}

Priors are designed to capture our existing knowledge of these parameters. Since we lack information on the possible values of these parameters, I choose diffuse priors with large variances. However, it is worth noting that the likelihood function has flat regions that can not be distinguished using only the data. Specifically, it is equally likely for the model to assign a large positive number (e.g., $100$) or a small negative number (e.g., $-100$) to a country's barrier level, as we did not provide information to the model about whether a positive number signifies a higher barrier level or a negative number does. To overcome this issue, I set semi-informative priors for two countries in the sample: Russia's barrier level is positive while Singapore's is negative. It tells the model that a positive number entails a higher barrier level than a negative one, as Russia has a higher level of barriers than Singapore to U.S. firms, according to anecdotal evidence.

Next, using the priors, I sample the augmented data $U_{ijt}^*$ from its conditional distribution:
\begin{align*}
	p(U_{ijt}^* | U, \Theta, \alpha^E, \alpha^R) = \begin{cases}
		\mathcal{N}_{(-\infty, \alpha_{ij}^R]}(\theta_{jt}, 1) \quad &\text{if} \quad U_{ijt} = 1 \\
		\mathcal{N}_{(\alpha_{ij}^R, \alpha_{ij}^E]}(\theta_{jt}, 1) \quad &\text{if} \quad U_{ijt} = 2\\
		\mathcal{N}_{( \alpha_{ij}^E, \infty)}(\theta_{jt}, 1) \quad &\text{if} \quad U_{ijt} = 3 \\
	\end{cases}
\end{align*}
the notation $\mathcal{N}_{(a,b)}$ denotes a truncated normal distribution on the support of $(a,b)$. This is the standard data augmentation step for models with latent variables \autocite{albert1993bayesian}.

After sampling the latent barrier $U_{ijt}^*$, I sample the reporting cutoffs $\alpha_{ij}^R$ from its posterior distribution.
\begin{align*}
	f(\alpha_{ij}^R | U, U^*, \Theta, \alpha^E) &= 	L(\Theta, \alpha^E, \alpha^R | U, U^*)  \cdot p(\alpha_{ij}^R | \alpha_{ij}^E) \\ 
	&\propto \prod_{t = 1}^T [I(U_{ijt} = 1, U^*_{ijt} \leq  \alpha_{ij}^R) + I(U_{ijt} = 2,   \alpha_{ij}^R  < U^*_{ijt} \leq  \alpha_{ij}^E)] \cdot p(\alpha_{ij}^R | \alpha_{ij}^E)
\end{align*}

Note that the product of $T$ indicator functions truncate the prior distribution $p(\alpha_{ij}^R | \alpha_{ij}^E)$ and only allows it to have positive densities over the interval $[\max(U_{ijt}^* | U_{ijt} = 1), \\ \min(U_{ijt}^* | U_{ijt} = 2))$. Analogously, the posterior distribution of $\alpha_{ij}^E$ is proportional to the prior distribution $p(\alpha_{ij}^E | \alpha_{ij}^R)$ truncated on the interval $[\max(U_{ijt}^* | U_{ijt} = 2), \min(U_{ijt}^* | U_{ijt} = 3))$. 

To sample the barrier level $\theta_{jt}$, we need to define its evolution probability. That is, how a country's barrier level fluctuates across years. In this paper, I model the state transition as a random walk:
\begin{align*}
	\theta_{j,t} = \theta_{j,t-1} + \delta_{jt}
\end{align*}
where $\delta_{jt}$ follows a normal distribution $\mathcal{N}(0, \sigma^2)$. I fix the variance term $\sigma^2$ a priori as $1$ for identification. However, it is worth noting that the variance term $\sigma$ smoothes the barrier level across years because the degree to which an estimated $\theta_{jt}$ shrinks back to the prior mean is inversely proportional to the variance of the disturbance term $\delta_{jt}$. For example, if a country's barrier level is completely independent of one another across years, or in other words, the barrier is not sticky at all, the disturbance variance would be infinity. Thus, fixing it as $1$ assumes a sticky regulatory barrier.

Finally, we sample the quantity of interest $\theta_{jt}$ from the full conditional posterior distribution $f(\Theta | \alpha^R, \alpha^E, U, U^*)$. A naive Gibbs Sampler approach requires sampling $\theta_{jt}$ conditional on $\theta_{j,t-1}$, which fails to incorporate the sticky evolution process and is hence inefficient. Literature has demonstrated that sampling highly correlated parameters with Gibbs Sampler can also lead to difficult-to-converge chains \autocite{martin2002dynamic}. To improve estimation, I adopt the Kalman filter to sample the entire times series at a time rather than sampling from the component by component conditional distributions (i.e., sample the entire time series $\theta_{j,k}$, $k \in \{1,2,\cdots, T\}$ instead of each $\theta_{jt}$ at a time). For ease of exposition, I omit the description of the forward-filtering and backward-sampling procedure.

One major weakness of this approach is that it cannot provide a regulatory barrier estimate for the U.S. because firms that file 10-K forms are U.S. firms. The U.S. is the home country for these firms, while any other country in which they operate is the host country. By definition, these firms must enter the U.S. market. And it is reasonable to argue that the data generating process is very different for the home and host countries. For these reasons, I exclude the U.S. from my analyses.

\section{Data and Results}

\subsection{Data}

My primary data source is the 10-K forms published by the U.S. Security and Exchange Commission (SEC). Since the model cannot distinguish between 1) a firm does not seek market entry and 2) the regulatory barrier is too restrictive for a firm to enter, the confounding effect of preference undermines the interpretability of the results. Although the proposed model accounts for \textit{time-invariant} unobserved dyadic heterogeneity (e.g., an energy company always prefers natural resource abundant countries), yearly fluctuations in global economic/political conditions and firms' financial performance may still bias my results. I partially circumvent this thorny issue by excluding 1) countries that host very few U.S. firms and 2) firms that have minimal international commercial activities. In essence, I aim to construct a sample in which firms' preferences can be viewed as constant so that any variation in firm behaviors must be due to the changing barrier levels. 

As a result, I adopt the following exclusion criteria: 
\begin{itemize}
	\item I exclude countries that more than 95\% of firms in my sample never entered. These are mostly African countries.
	\item I only include the top $1500$ firms that are most active in the international market between 2006 and 2015. A firm is deemed more active if it consistently operates in more countries.
\end{itemize}

The period of my focus is between 2006 and 2015. The 10-K database used in this paper is compiled by \textcite{loughran2016textual}, which covers 10-K reports from 1993. At the time of download in 2019, I excluded years after 2015 to avoid potential backfiling issues. Years before 2006 are also excluded to reduce computational complexity, as information from that period is relatively outdated.

After cleaning, I have a sample of 853 firms and 40 countries. Because I only keep firms that are consistently in the sample between 2006 and 2015, the number of firms drops from 1500 to 853.

\subsection{Results}

\begin{figure}[h!]
    \centering
    \includegraphics[scale=0.82]{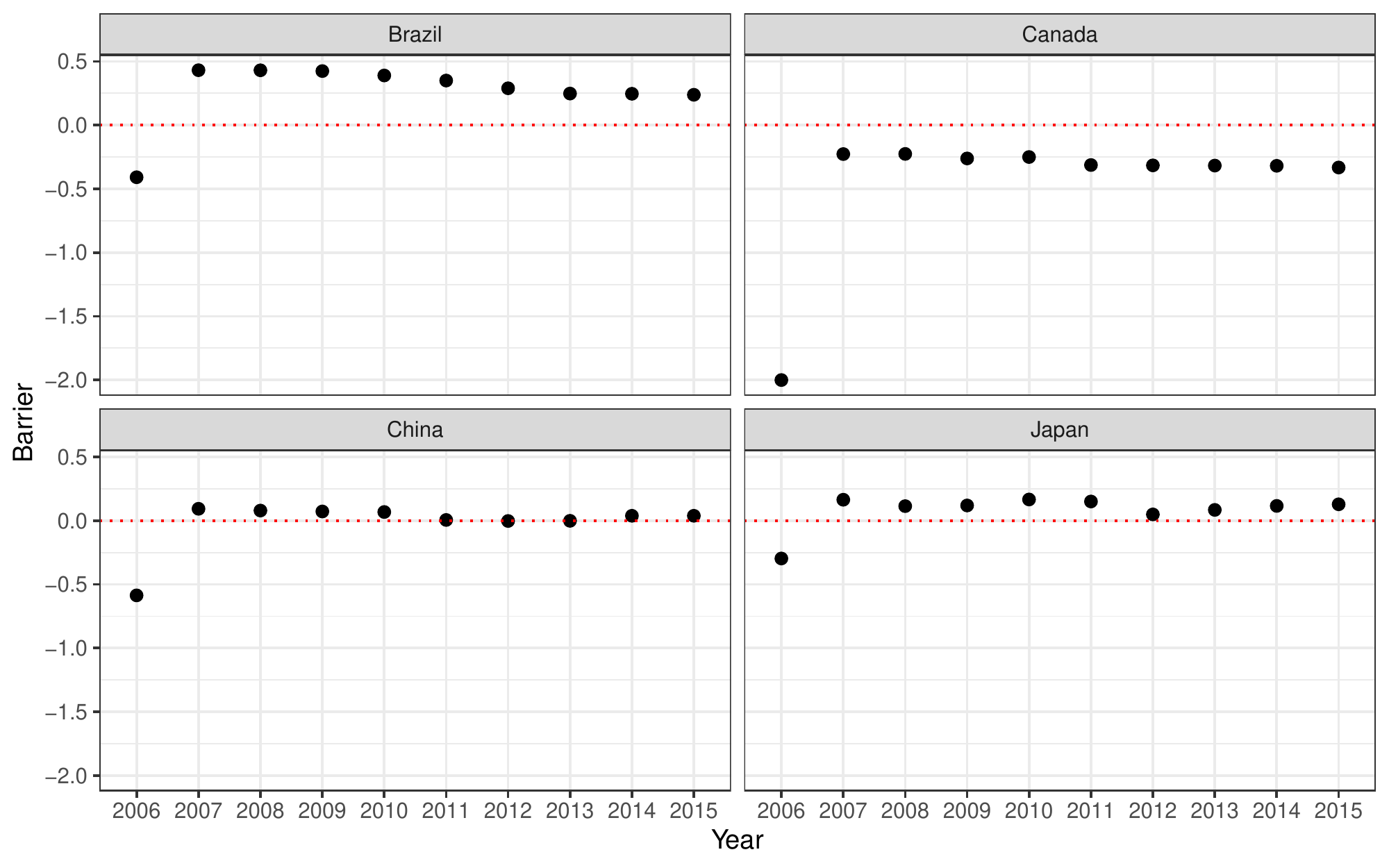}
    \caption{Temporal Change in Regulatory Barrier (Brazil, Canada, China, Japan)}
    \label{fig:comp}
\end{figure}

\hyperref[fig:comp]{Figure \ref{fig:comp}} presents the temporal change in the estimated barrier level for four countries (Brazil, Canada, China, Japan); however, readers can find the entire list of estimated barrier levels in the appendix. Since the four chosen countries have close economic connections with the U.S., examining their estimated barrier level is a preliminary test for the estimates' plausibility. 

Among the four countries, Canada has the lowest barrier level consistently. Recall that the estimates draw information from the U.S. firms' entry and reporting decisions. In the case of Canada, it demonstrates that more U.S. firms enter Canada, but fewer of them report barriers when compared with the three other countries. It is worth noting that the number of U.S. firms that report encountering barriers in Canada may still exceed the other three countries. Yet, Canada's barrier level is still estimated to be lower because the number of firms operating in Canada can be much higher than in other countries. 

Compared to the other three countries, China also displays a relatively low barrier level, which is counter-intuitive, as many U.S. firms report encountering barriers in China in my datasets. Again, this result should be driven by the large number of U.S. firms operating in China. However, the estimated difference in barrier level between China and Canada is still considerable, even though both countries host many U.S. firms. Thus, I believe that the model aggregates the firms' entry and reporting information in a consistent and reasonable manner. 

The estimates show a sharp jump in the barrier level across all countries from 2006 to 2007, suggesting a global shock in 2007. Among the four countries, the jump is the largest for Canada, followed by Brazil, China, and Japan. It is difficult to pinpoint the cause of the jump using only the information from the dataset. However, I offer some suggestive evidence regarding the possible causes. In 2006, there were 2,305 incidences of barrier reporting, while that number increased to 2,636 in 2007, a 14.4\% increase. A plausible reason could be the 2007-2008 global financial crisis which should impact the regulatory barrier levels globally. It appears that more U.S. firms had financial difficulties in 2007 than in 2006. For example, here is a list of statements on bankruptcy in the 2007 10-Ks, whose number increases by 20\% from 2006 to 2007:
\begin{itemize}
    \item  ``Some of our current and former international customers, particularly automobile manufacturers in Europe and Japan, were reluctant to do business with us while we underwent chapter 11 bankruptcy.''
    \item ``The proposed transaction is subject to approval by the United States Bankruptcy Court, receipt of required regulatory approvals, finalizing the definitive purchase agreement for Akzo Nobel's Crystex.''
\end{itemize}
Still, it is worth emphasizing that more systemic analyses are required to understand the financial crisis's effect on the observed barrier level jump.

\begin{figure}[h!]
    \centering
    \includegraphics[scale=0.82]{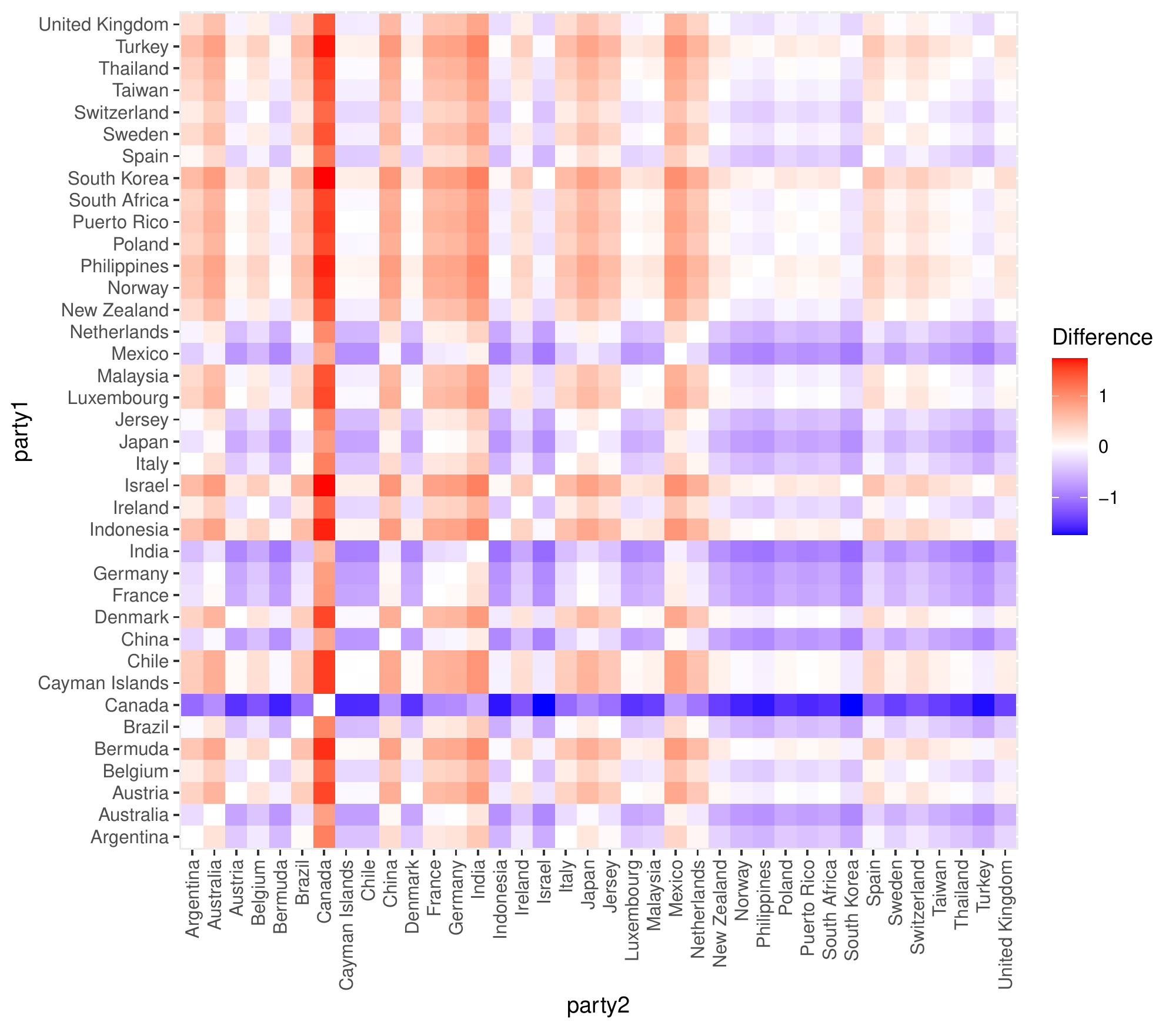}
    \caption{Pairwise Comparison of Countries' Average Barrier Level}
    \label{fig:heatmap}
\end{figure}

\hyperref[fig:heatmap]{Figure \ref{fig:heatmap}} presents a cross-country comparison of the estimated regulatory barrier level. Each cell in the heat map is the barrier difference between a pair of countries labeled by the axis ticks. The country's barrier level is the average of its levels across years. Specifically, a negative value in the cell signifies that the country represented by the Y axis (i.e., party 1) has a lower barrier level than that on the X axis (i.e., party 2). It can be observed that the row of Canada is the bluest among all, which shows that Canada has the lowest average barrier level among all countries.

Similarly, India, Germany, and France also display significantly lower barrier levels than other countries on average. On the other hand, Philippines, Norway, and New Zealand have relatively higher barriers than other countries on average. On the other hand, China has a medium level of barriers compared with the rest of the world.

\hyperref[fig:comp]{Figure \ref{fig:comp}} and \hyperref[fig:heatmap]{Figure \ref{fig:heatmap}} serve as preliminary validation tests of the estimated barrier. However, more rigorous results are needed to establish the accuracy and consistency of the proposed index. Thus, I present several statistical analyses in the next section, which compare and contrast my index with other popular measurements in the field.

\section{Validation}

\subsection{Special Trade Concerns (STCs)}

\textcite{gulotty_narrowing_2018} has shown in his book that the special trade concerns (STCs) data, and more specifically, the technical barrier to trade (TBT) data collected by the World Trade Organization (WTO) can inform researchers of the regulatory barrier levels of major economies in the world. 

As I briefly explained in the introduction section, my proposed index can be superior to the STC-TBT data in two major aspects:
\begin{itemize}
    \item \textcite{gulotty_narrowing_2018} noted that: ``the choice to raise a foreign regulation as an STC is as much a political process as the choice to impose the regulation in the ﬁrst place.'' Thus, the STC-TBT data is likely to be heavily influenced by international politics. This concern is challenging to eliminate but can be fatal for researchers who study the correlation between regulatory barriers and international relations. However, my estimated barrier is less susceptible to such a concern as it is unlikely that international politics may shape an individual firm's decision to report in their annual reports.
    \item It is known that STC-TBT data tend to target larger markets, as governments need to balance the cost of filing an STC complaint and its benefit on the domestic economy \autocite{fontagne2018let,gulotty_narrowing_2018}. Therefore, the observed STC-TBT report distribution is heavily skewed along the market size dimension: countries with larger market sizes are more likely to be included in STC reports than their smaller counterparts, even if they have the same barrier level. Admittedly, my index cannot eliminate the contaminating effect of market sizes. Still, by accounting for time-invariant dyadic-specific confounders, the proposed index should address the concern more satisfyingly.
\end{itemize}

\begin{figure}[h!]
    \centering
    \begin{subfigure}[b]{\linewidth}
        \centering
        \includegraphics[scale=0.65]{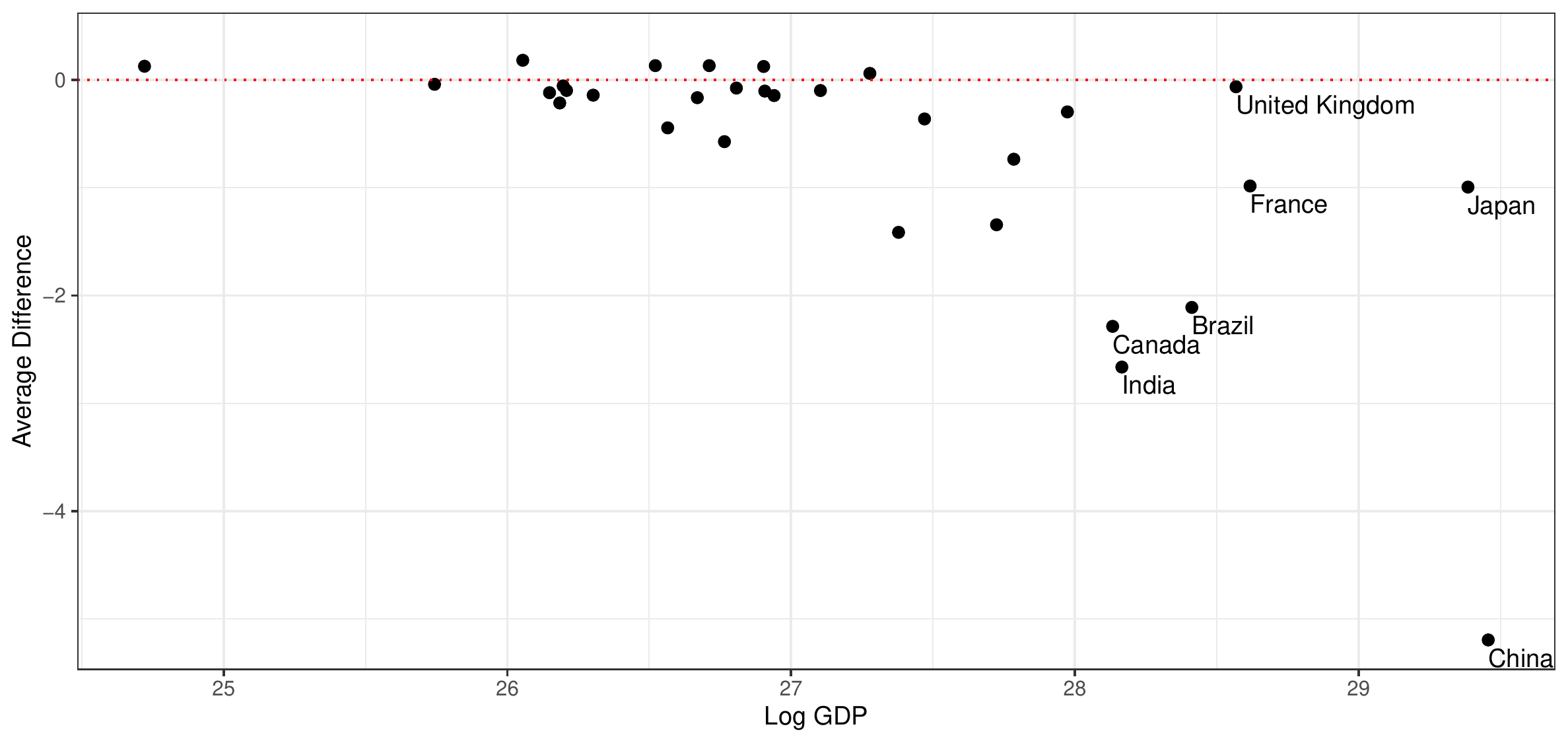}
        \caption{Comparison with STCs count}
        \label{stc:gdp}
    \end{subfigure}
    \begin{subfigure}[b]{\linewidth}
        \centering
        \includegraphics[scale=0.65]{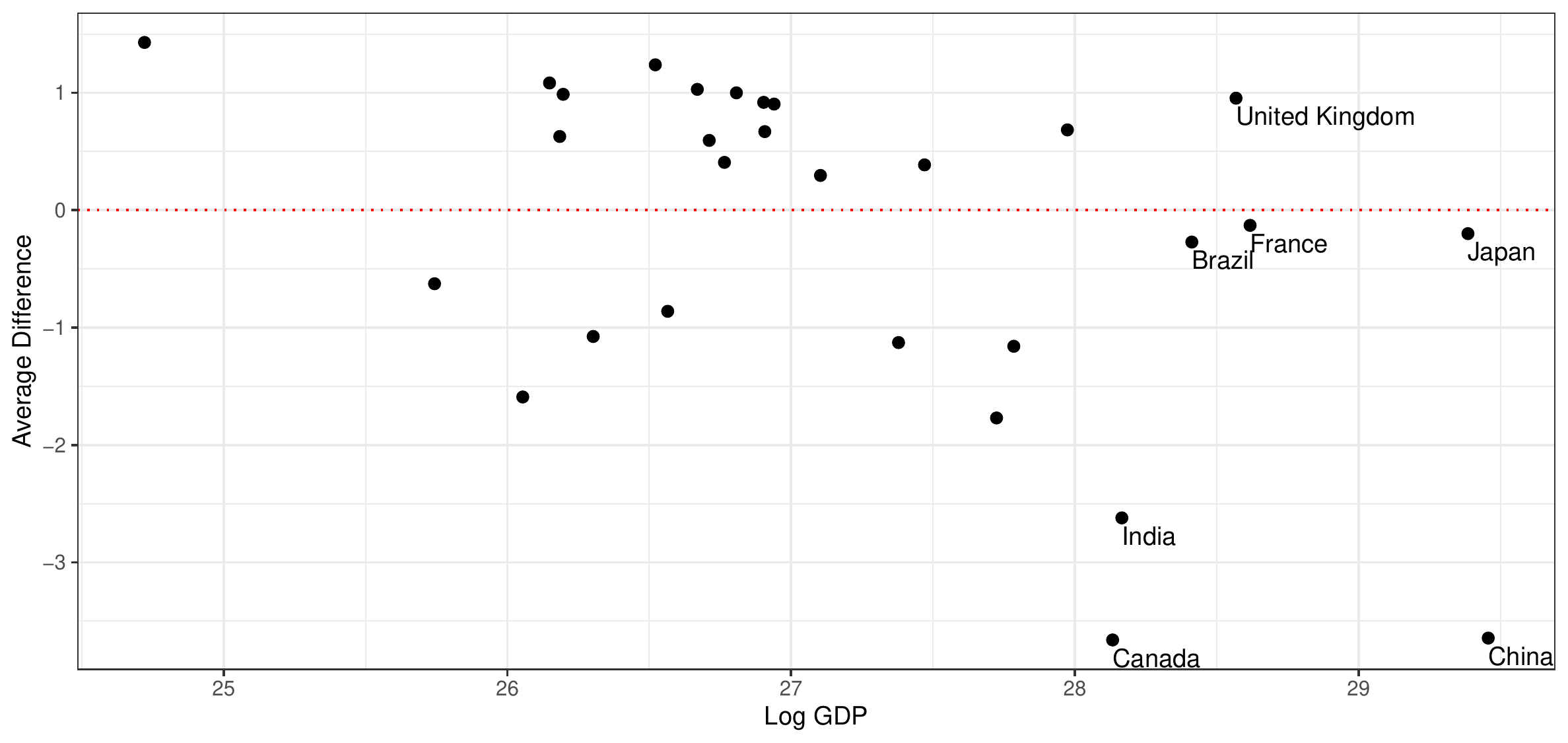}
        \caption{Comparison with OECD FDI Regulatory Restrictiveness Index}
        \label{fdi:gdp}
    \end{subfigure}
    \label{fig:gdp_barrier}
\end{figure}

I conduct two validation analyses to compare the STC-TBT data and my proposed index. First, I normalize the STCs count that a country is subject to and its estimated barrier by subtracting their mean and dividing by the standard deviation. Then, I take the difference between these two data at the country level and plot the difference against the market size of the countries, measured by GDP. The result is presented in \hyperref[stc:gdp]{Figure \ref{stc:gdp}}. The horizontal axis shows the log GDP of each country: countries with a large GDP are placed more on the right. The vertical axis is the normalized difference between the STCs count and the estimated barrier level: a negative value shows that my estimated barrier level is lower than the normalized STCs count. We observe that the difference is more pronounced in countries with large market sizes. More specifically, the barrier level measured by my index is consistently lower than that measured by the STCs count among countries with large market sizes. On the contrary, the two measurements align quite well among small countries. These findings are consistent with the observation that the STC-TBT data often inflate the barrier level of large economies. The proposed index suffers less from such a weakness.

Next, I regress my estimated barrier level on the count of STC-TBT reports filed against each country between 2006 and 2015. Column (1), (2), and (3) in \hyperref[reg:stcfdi]{Table \ref{reg:stcfdi}} display the regression results. The bi-variate regression between the proposed index and the STCs count returns a negative coefficient. However, the coefficient becomes positive after accounting for country, year fixed effects, and GDP/GDP per capita. Although the positive correlation fails to achieve statistical significance, the results nonetheless corroborate my claims on the contaminating effect of market sizes and the weakness of the STC-TBT data.

\begin{table}[h!]
\centering
\input{fig/stc_fdi_reg.tex}
\caption{Regression Analyses of STCs, FDI Restrictiveness Index and the Estimated Barrier}
\label{reg:stcfdi}
\end{table}

\subsection{OECD FDI Regulatory Restrictiveness Index}

Next, I compare my barrier index with the popular OECD FDI Regulatory Restrictiveness Index \autocite{koyama2006oecd,kalinova2010oecd}. The index covers four types of regulatory measures: (1) foreign equity restrictions, (2) screening and prior approval requirements, (3) rules for key personnel, and (4) other restrictions on the operation of foreign enterprises. In this analysis, I use the aggregate index at the country-year level.

First, I plot log GDP against the normalized difference between FDI regulatory restrictiveness index and the estimated barrier (\hyperref[fdi:gdp]{Figure \ref{fdi:gdp}}). There is a slight negative relationship between the average difference and log GDP, similar to what we observe in the STC-TBT case. China, India, and Canada are still among the countries that enjoy a considerable negative difference. That is, my proposed index assigns a significantly lower barrier level than the FDI regulatory restrictiveness index. However, the observed negative correlation is much less pronounced than it is in the STC-TBT case, which suggests that the FDI regulatory restrictiveness index captures issues that differ from what STC-TBT capture.

I proceed to regress my estimated barrier index on the FDI regulatory restrictiveness index and report the results in column (4), (5), and (6) in \hyperref[reg:stcfdi]{Table \ref{reg:stcfdi}}. However, it is surprising that the coefficients of the FDI restrictiveness index are consistently negative, suggesting a strong negative correlation between my proposed barrier estimates and the restrictiveness index. The negative relationship persists even after accounting for market sizes and country/year idiosyncrasies. It shows that countries with a higher FDI regulatory restrictiveness index are often associated with a lower regulatory barrier level, per my proposed estimates. However, readers need to be cautious when interpreting this seemingly contradictory result, as the two indices may simply reflect different aspects of regulatory barriers. The FDI regulatory restrictiveness index focuses on the regulatory restrictions that \textit{only affect} foreign firms. At the same time, my proposed barrier aims to measure regulation that affects both domestic and foreign firms but may constitute hidden obstacles for foreign firms in practice. The results could demonstrate that countries that use domestic regulations as hidden barriers can adopt less restrictive FDI-targeting measures, as blatant restrictions on foreign ownership may imply significant political costs both domestically and internationally \autocite{kono2008democracy,esberg2021covert}.

\subsection{Trade, FDI, and Democracy}

Finally, I examine the relationship between my proposed barrier estimates and the U.S. trade volume, FDI flow, FDI stock, and regime types. Intuitively, a higher barrier should be correlated with lower trade flow, lower FDI flow, and lower FDI stock. I also revisit the classic debate between regime types and regulatory barrier \autocite{milner2005move,kono2008democracy,pandya2014democratization}.

\begin{figure}[h!]
    \centering
    \includegraphics[scale=0.65]{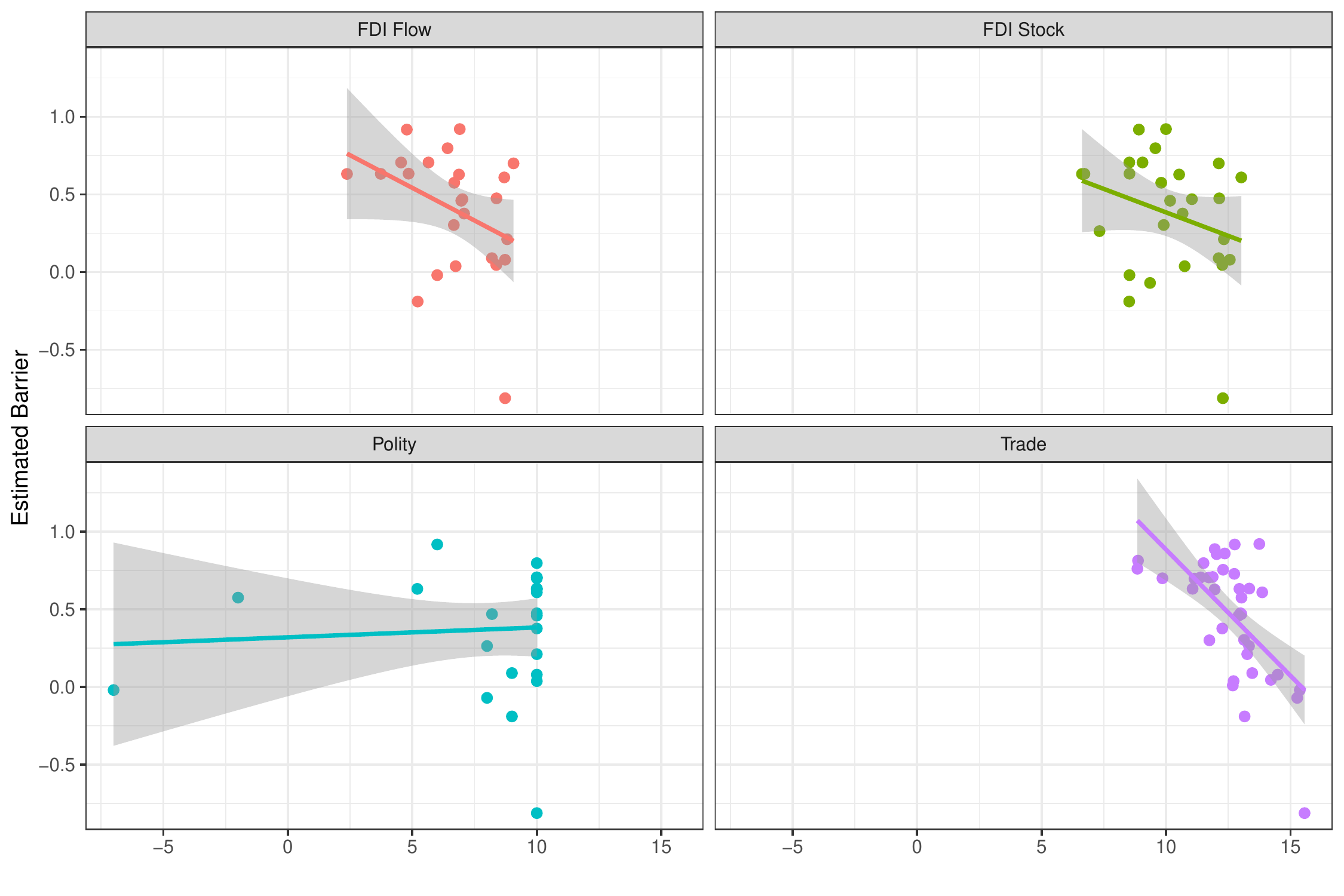}
    \caption{Correlation with Major Indicators (Polity, Trade, FDI Flow, FDI Stock)}
    \label{fig:corr}
\end{figure}

I visualize the results in \hyperref[fig:corr]{Figure \ref{fig:corr}}. Each panel represents the overall correlation between my proposed barrier estimate and trade flow, FDI flow, FDI stock, and the Polity 2 score. It can be clearly observed that a country with a higher barrier level is associated with 1) lower trade volume with the U.S., 2) receives less FDI from the U.S., and 3) has lower FDI stock from the U.S.. However, the correlation between the Polity 2 score and the estimated barrier level is very weak. These results lend further credibility to my proposed barrier estimates.

\section{Conclusions}
This paper offers a novel measurement of the elusive quantity: regulatory barrier, contributing to the empirical literature on the political economy of regulation and regulatory barrier. I leverage information in the annual reports of U.S. public firms (i.e., 10-Ks forms) to best address two major concerns when measuring regulatory barriers: (1) the confounding effect of international politics and (2) the bias caused by market size. I show in a series of validation analyses that the proposed barrier index shows patterns consistent with our existing knowledge of regulatory barriers. Moreover, the new barrier estimates display signs of better addressing the problems mentioned in the previous text than the existing measurements. 

The proposed barrier index serves as an additional measurement of regulatory barriers, which may be superior to existing ones in some research contexts. I intend not to claim that my estimate precisely captures the concept of regulatory barrier and hence is the ``best'' one. However, my goal is to shed light on new aspects of regulatory barrier that evades researchers' attention due to their hard-to-observe nature. There are obvious weaknesses in both the information sources and my models. Nonetheless, my estimates offer new insights into this important political and economic phenomenon.

The paper also contributes to international political economy and international relations by proposing the firms' annual reports as a new source of information for further research. Since high-quality text data has become increasingly important for empirical research in political science, the rich information contained in the 10-Ks firms deserves more scholarly attention in the future. 

\clearpage

\section{Appendix}

\subsection{The Full List}

\singlespacing
\footnotesize
I list the estimated barrier levels for all 40 countries below.
	\begin{tabularx}{\linewidth}{XXXXXXXXXXX}
	\toprule
Country & 2006 & 2007 & 2008 & 2009 & 2010 & 2011 & 2012 & 2013 & 2014 & 2015 \\ 
 \midrule
 \endhead
Singapore & -0.18 & 0.68 & 0.70 & 0.71 & 0.68 & 0.67 & 0.62 & 0.61 & 0.63 & 0.64 \\ 
Australia & -1.12 & 0.24 & 0.18 & 0.17 & 0.16 & 0.17 & 0.16 & 0.15 & 0.13 & 0.13 \\ 
Austria & 0.12 & 0.81 & 0.78 & 0.77 & 0.78 & 0.76 & 0.73 & 0.76 & 0.76 & 0.79 \\ 
Belgium & -0.23 & 0.62 & 0.55 & 0.58 & 0.56 & 0.54 & 0.52 & 0.51 & 0.51 & 0.54 \\ 
Bermuda & 0.32 & 0.88 & 0.90 & 0.86 & 0.85 & 0.84 & 0.87 & 0.89 & 0.86 & 0.88 \\ 
Brazil & -0.41 & 0.43 & 0.43 & 0.42 & 0.39 & 0.35 & 0.29 & 0.25 & 0.25 & 0.24 \\ 
Canada & -5.56 & -0.23 & -0.23 & -0.26 & -0.25 & -0.31 & -0.32 & -0.32 & -0.32 & -0.33 \\ 
Cayman Islands & 0.03 & 0.91 & 0.85 & 0.85 & 0.87 & 0.83 & 0.78 & 0.83 & 0.83 & 0.83 \\ 
Chile & 0.09 & 0.95 & 0.88 & 0.91 & 0.86 & 0.84 & 0.79 & 0.74 & 0.73 & 0.76 \\ 
China & -0.59 & 0.09 & 0.08 & 0.07 & 0.07 & 0.01 & -0.00 & -0.00 & 0.04 & 0.04 \\ 
Denmark & 0.09 & 0.84 & 0.78 & 0.80 & 0.76 & 0.76 & 0.75 & 0.77 & 0.78 & 0.74 \\ 
France & -0.51 & 0.17 & 0.15 & 0.16 & 0.16 & 0.13 & 0.15 & 0.16 & 0.16 & 0.17 \\ 
Germany & -0.47 & 0.14 & 0.13 & 0.13 & 0.13 & 0.10 & 0.08 & 0.08 & 0.06 & 0.08 \\ 
India & -0.62 & -0.09 & -0.11 & -0.12 & -0.14 & -0.13 & -0.16 & -0.16 & -0.17 & -0.19 \\ 
Indonesia & 0.29 & 1.00 & 0.95 & 0.94 & 0.97 & 0.92 & 0.91 & 0.87 & 0.86 & 0.89 \\ 
Ireland & 0.03 & 0.62 & 0.55 & 0.52 & 0.51 & 0.51 & 0.43 & 0.43 & 0.48 & 0.50 \\ 
Israel & 0.34 & 0.99 & 1.01 & 1.01 & 1.02 & 0.97 & 0.98 & 0.99 & 0.95 & 0.92 \\ 
Italy & -0.11 & 0.38 & 0.37 & 0.37 & 0.38 & 0.39 & 0.31 & 0.29 & 0.30 & 0.33 \\ 
Japan & -0.30 & 0.16 & 0.11 & 0.12 & 0.17 & 0.15 & 0.05 & 0.08 & 0.12 & 0.13 \\ 
Jersey & -0.14 & 0.26 & 0.22 & 0.24 & 0.30 & 0.33 & 0.33 & 0.35 & 0.34 & 0.35 \\ 
Luxembourg & 0.39 & 0.87 & 0.86 & 0.82 & 0.79 & 0.72 & 0.68 & 0.64 & 0.63 & 0.60 \\ 
Malaysia & 0.32 & 0.75 & 0.71 & 0.70 & 0.69 & 0.67 & 0.65 & 0.62 & 0.59 & 0.61 \\ 
Mexico & -0.37 & -0.05 & -0.02 & -0.01 & -0.03 & -0.06 & -0.03 & -0.04 & -0.03 & -0.06 \\ 
Netherlands & -0.02 & 0.26 & 0.26 & 0.28 & 0.26 & 0.24 & 0.20 & 0.21 & 0.21 & 0.20 \\ 
New Zealand & 0.40 & 0.73 & 0.67 & 0.71 & 0.68 & 0.65 & 0.62 & 0.62 & 0.61 & 0.62 \\ 
Norway & 0.51 & 0.92 & 0.85 & 0.88 & 0.85 & 0.85 & 0.80 & 0.81 & 0.76 & 0.74 \\ 
Philippines & 0.59 & 0.92 & 0.88 & 0.92 & 0.93 & 0.88 & 0.90 & 0.85 & 0.84 & 0.85 \\ 
Poland & 0.50 & 0.80 & 0.75 & 0.75 & 0.72 & 0.72 & 0.69 & 0.67 & 0.66 & 0.70 \\ 
Puerto Rico & 0.53 & 0.80 & 0.81 & 0.81 & 0.82 & 0.81 & 0.79 & 0.74 & 0.75 & 0.76 \\ 
United Kingdom & 0.60 & 0.82 & 0.68 & 0.65 & 0.64 & 0.61 & 0.57 & 0.54 & 0.50 & 0.47 \\ 
Argentina & 0.23 & 0.42 & 0.38 & 0.36 & 0.34 & 0.30 & 0.25 & 0.24 & 0.24 & 0.25 \\ 
South Africa & 0.65 & 0.81 & 0.77 & 0.77 & 0.75 & 0.71 & 0.68 & 0.67 & 0.63 & 0.65 \\ 
South Korea & 0.85 & 1.01 & 1.01 & 1.00 & 0.96 & 0.93 & 0.86 & 0.86 & 0.87 & 0.87 \\ 
Spain & 0.31 & 0.42 & 0.40 & 0.43 & 0.40 & 0.41 & 0.33 & 0.33 & 0.34 & 0.40 \\ 
Sweden & 0.58 & 0.70 & 0.62 & 0.64 & 0.66 & 0.63 & 0.62 & 0.62 & 0.62 & 0.59 \\ 
Switzerland & 0.43 & 0.55 & 0.52 & 0.51 & 0.50 & 0.46 & 0.45 & 0.46 & 0.43 & 0.44 \\ 
Taiwan & 0.61 & 0.66 & 0.67 & 0.68 & 0.67 & 0.64 & 0.60 & 0.59 & 0.60 & 0.62 \\ 
Thailand & 0.78 & 0.81 & 0.78 & 0.77 & 0.76 & 0.75 & 0.67 & 0.64 & 0.66 & 0.66 \\ 
Turkey & 1.02 & 1.03 & 0.98 & 0.93 & 0.89 & 0.86 & 0.79 & 0.81 & 0.79 & 0.78 \\ 
Russia & 0.03 & 0.05 & 0.03 & 0.01 & -0.00 & 0.01 & 0.02 & -0.02 & 0.02 & -0.06 \\ 
  \bottomrule
\end{tabularx}
\clearpage

\doublespacing
\normalsize

\subsection{Regression Analyses of Trade, FDI, and Regime Type}

\begin{table}[h!]
\resizebox{\textwidth}{!}{
\input{fig/economic_reg.tex}
}
\caption{Regression Analyses of Trade, FDI, and Regime Type}
\label{reg:indicator}
\end{table}

\hyperref[reg:indicator]{Table \ref{reg:indicator}} presents the regression results of the same exercise. Column 1 to 4 reports the bi-variate regression results that are visualized in \hyperref[fig:corr]{Figure \ref{fig:corr}}. In column 5 to 8, I include country/year fixed effects and economic variables as covariates. Interestingly, the coefficient of Polity 2 flips and becomes statistically significant, suggesting that democracies are associated with lower estimated barrier. However, the coefficients of the three other economic variables fail to achieve statistical significance after including the fixed effects and covariates.

\end{document}

%% file: fig/stc_fdi_reg.tex
\begin{tabular}{lcccccc}
\tabularnewline\toprule\toprule
&\multicolumn{6}{c}{Estimated Barrier}\\
Model:&(1) & (2) & (3) & (4) & (5) & (6)\\
\midrule
&  & & & & & \\
STCs Count&-0.0967$^{***}$&0.0375&0.0386&  &  &  \\
  &(0.0230)&(0.0349)&(0.0351)&  &  &  \\
FDI Restrictiveness Index&  &  &  &-0.3421&-4.702$^{***}$&-6.371$^{***}$\\
  &  &  &  &(0.3014)&(0.9352)&(1.541)\\
Log GDP&  &  &0.4774&  &  &-0.5753$^{**}$\\
  &  &  &(0.6072)&  &  &(0.2855)\\
Log GDP pc&  &  &-0.4406&  &  &0.4051\\
  &  &  &(0.5846)&  &  &(0.2491)\\
\midrule
Year&&$\checkmark$&$\checkmark$&&$\checkmark$&$\checkmark$\\
Country&&$\checkmark$&$\checkmark$&&$\checkmark$&$\checkmark$\\
\midrule
\emph{Fit statistics}&  & & & & & \\
Observations& 400&400&310&231&231&203\\
R$^2$ & 0.04266&0.72868&0.69764&0.0056&0.69117&0.68889\\
\bottomrule\bottomrule
\multicolumn{7}{l}{\emph{Signif. Codes: ***: 0.01, **: 0.05, *: 0.1}}\\
\end{tabular}

%% file: fig/economic_reg.tex
\begin{tabular}{lcccccccc}
\tabularnewline\toprule\toprule
&\multicolumn{8}{c}{Estimated Barrier}\\
Model:&(1) & (2) & (3) & (4) & (5) & (6) & (7) & (8)\\
\midrule
Log Trade&-0.1268$^{***}$& & &  &-0.0077&&&  \\
  &(0.0093)&&&  &(0.0125)&&&  \\
Log FDI Flow& &-0.0628$^{***}$&&  &&-0.0096&&  \\
  &&(0.0206)&&  &&(0.0165)&&  \\
Log FDI Stock&&&-0.0464$^{***}$&  &&&-0.0323&  \\
  &&&(0.0163)&  &&&(0.0397)&  \\
Polity2&  &  &  &0.0043&  &  &  &-0.0260$^{***}$\\
  &  &  &  &(0.0052)&  &  &  &(0.0067)\\
Log GDP&  &  &  &  &0.5403&0.2625&0.3575&0.5571\\
  &  &  &  &  &(0.5702)&(0.7483)&(0.6548)&(0.5560)\\
Log GDP pc&  &  &  &  &-0.5018&-0.1287&-0.0892&-0.5041\\
  &  &  &  &  &(0.6377)&(0.8399)&(0.7496)&(0.6230)\\
\midrule
\emph{Fixed-effects}&  & & & & & & & \\
Country&&&&&$\checkmark$&$\checkmark$&$\checkmark$&$\checkmark$\\
Year&&&&&$\checkmark$&$\checkmark$&$\checkmark$&$\checkmark$\\
\midrule
Observations& 760&217&268&620&620&189&238&620\\
R$^2$ & 0.19626&0.04152&0.02957&0.00114&0.69519&0.65353&0.67206&0.69698\\
\bottomrule\bottomrule
\multicolumn{9}{l}{\emph{Signif. Codes: ***: 0.01, **: 0.05, *: 0.1}}\\
\end{tabular}

%% file: ref.bib
@book{gulotty_narrowing_2018,
title={Narrowing the Channel: The Politics of Regulatory Protection in International Trade},
author={Gulotty, Robert},
year={2020},
publisher={University of Chicago Press}
}

@book{buthe_new_2011,
	title = {The New Global Rulers: The Privatization of Regulation in the World Economy},
	shorttitle = {The New Global Rulers},
	publisher = {Princeton University Press},
	author = {Büthe, Tim and Mattli, Walter},
	year = {2011}
}

@article{kennard2017firms,
title={The Enemy of My Enemy: When Firms Support Climate Change Regulation},
author={Kennard, Amanda},
journal={International Organization},
volume={74},
number={2},
pages={187--221},
year={2020},
publisher={Cambridge University Press}
}

@book{carpenter2013preventing,
	title={Preventing regulatory capture: Special interest influence and how to limit it},
	author={Carpenter, Daniel and Moss, David A},
	year={2013},
	publisher={Cambridge University Press}
}

@book{davis2012adjudicate,
	title={Why adjudicate?: enforcing trade rules in the WTO},
	author={Davis, Christina L},
	year={2012},
	publisher={Princeton University Press}
}

@article{cooley2019barrier,
	title={Estimating Policy Barriers to Trade},
	author={Cooley, Brendan},
	journal={working paper},
	year={2019}
}

@article{hollyer2014measuring,
	title={Measuring transparency},
	author={Hollyer, James R and Rosendorff, B Peter and Vreeland, James Raymond},
	journal={Political analysis},
	volume={22},
	number={4},
	pages={413--434},
	year={2014},
	publisher={Cambridge University Press}
}

@article{clinton2004statistical,
	title={The statistical analysis of roll call data},
	author={Clinton, Joshua and Jackman, Simon and Rivers, Douglas},
	journal={American Political Science Review},
	volume={98},
	number={2},
	pages={355--370},
	year={2004},
	publisher={Cambridge University Press}
}

@article{perlman2019,
	title={For Safety or Profit? How Science Serves the Strategic Interests of Private Actors},
	author={Perlman, Rebecca},
	journal={American Journal of Political Science},
	year={2019}
}

@article{campbell2014information,
	title={The information content of mandatory risk factor disclosures in corporate filings},
	author={Campbell, John L and Chen, Hsinchun and Dhaliwal, Dan S and Lu, Hsin-min and Steele, Logan B},
	journal={Review of Accounting Studies},
	volume={19},
	number={1},
	pages={396--455},
	year={2014},
	publisher={Springer}
}

@article{brown2018spillover,
	title={The spillover effect of SEC comment letters on qualitative corporate disclosure: Evidence from the risk factor disclosure},
	author={Brown, Stephen V and Tian, Xiaoli and Wu Tucker, Jennifer},
	journal={Contemporary Accounting Research},
	volume={35},
	number={2},
	pages={622--656},
	year={2018},
	publisher={Wiley Online Library}
}

@article{devlin2018bert,
	title={Bert: Pre-training of deep bidirectional transformers for language understanding},
	author={Devlin, Jacob and Chang, Ming-Wei and Lee, Kenton and Toutanova, Kristina},
	journal={arXiv},
	year={2018}
}

@article{bailey2017estimating,
	title={Estimating dynamic state preferences from United Nations voting data},
	author={Bailey, Michael A and Strezhnev, Anton and Voeten, Erik},
	journal={Journal of Conflict Resolution},
	volume={61},
	number={2},
	pages={430--456},
	year={2017},
	publisher={SAGE Publications Sage CA: Los Angeles, CA}
}

@article{martin2002dynamic,
	title={Dynamic ideal point estimation via Markov chain Monte Carlo for the US Supreme Court, 1953--1999},
	author={Martin, Andrew D and Quinn, Kevin M},
	journal={Political analysis},
	volume={10},
	number={2},
	pages={134--153},
	year={2002},
	publisher={Cambridge University Press}
}

@book{west2006bayesian,
	title={Bayesian forecasting and dynamic models},
	author={West, Mike and Harrison, Jeff},
	year={2006},
	publisher={Springer Science \& Business Media}
}

@article{albert1993bayesian,
	title={Bayesian analysis of binary and polychotomous response data},
	author={Albert, James H and Chib, Siddhartha},
	journal={Journal of the American statistical Association},
	volume={88},
	number={422},
	pages={669--679},
	year={1993},
	publisher={Taylor \& Francis}
}

@article{loughran2016textual,
	title={Textual analysis in accounting and finance: A survey},
	author={Loughran, Tim and McDonald, Bill},
	journal={Journal of Accounting Research},
	volume={54},
	number={4},
	pages={1187--1230},
	year={2016},
	publisher={Wiley Online Library}
}

@article{fontagne2018let,
  title={Let’s try next door: Technical Barriers to Trade and multi-destination firms},
  author={Fontagn{\'e}, Lionel and Orefice, Gianluca},
  journal={European Economic Review},
  volume={101},
  pages={643--663},
  year={2018},
  publisher={Elsevier}
}

@incollection{koyama2006oecd,
  title={OECD's FDI Regulatory Restrictiveness Index},
  author={Koyama, Takeshi and Golub, Stephen S and others},
  booktitle={OECD Economics Department Working Paper},
  year={2006},
  publisher={Citeseer}
}

@article{kalinova2010oecd,
  title={OECD's FDI restrictiveness index: 2010 update},
  author={Kalinova, Blanka and Palerm, Angel and Thomsen, Stephen},
  year={2010},
  publisher={OECD}
}

@article{kono2008democracy,
  title={Democracy and trade discrimination},
  author={Kono, Daniel Yuichi},
  journal={The Journal of Politics},
  volume={70},
  number={4},
  pages={942--955},
  year={2008},
  publisher={Cambridge University Press New York, USA}
}

@article{esberg2021covert,
  title={Covert Confiscation: How Governments Differ in Their Strategies of Expropriation},
  author={Esberg, Jane and Perlman, Rebecca},
  journal={Comparative Political Studies},
  pages={00104140221089650},
  year={2021},
  publisher={SAGE Publications Sage CA: Los Angeles, CA}
}

@article{milner2005move,
  title={Why the move to free trade? Democracy and trade policy in the developing countries},
  author={Milner, Helen V and Kubota, Keiko},
  journal={International organization},
  volume={59},
  number={1},
  pages={107--143},
  year={2005},
  publisher={Cambridge University Press}
}

@article{pandya2014democratization,
  title={Democratization and foreign direct investment liberalization, 1970--2000},
  author={Pandya, Sonal S},
  journal={International Studies Quarterly},
  volume={58},
  number={3},
  pages={475--488},
  year={2014},
  publisher={Blackwell Publishing Ltd Oxford, UK}
}

@article{martini2020backward,
  title={Backward-engineering trade protection: how to estimate worldwide industry-level trade barriers},
  author={Martini, Marco},
  year={2020},
  publisher={University of Zurich}
}

@article{mccubbins1984congressional,
  title={Congressional oversight overlooked: Police patrols versus fire alarms},
  author={McCubbins, Mathew D and Schwartz, Thomas},
  journal={American journal of political science},
  pages={165--179},
  year={1984},
  publisher={JSTOR}
}
